# Distortionless pulse transmission in valley photonic crystal slab waveguide

Fu-Long Shi[1,†], Yuan Cao[2,†], Xiao-Dong Chen[1,†], Jian-Wei Liu[1], Wen-Jie Chen[1], Min Chen[1,3,*] and Jian-Wen Dong[1,*]

[1]*School of Physics & State Key Laboratory of Optoelectronic Materials and Technologies, Sun Yat-sen University, Guangzhou 510275, China.*

[2]*Guangdong Provincial Key Laboratory of Optical Fiber Sensing and Communications, Institute of Photonics Technology, Jinan University, Guangzhou 510632, China*

3, *Department of Physics, College of Science, Shantou University, Shantou 510063, China.*

[†]These authors contributed equally to this work

[*]Corresponding author: stscm@mail.sysu.edu.cn & dongjwen@mail.sysu.edu.cn

## ABSTRACT

**Valley photonic crystal is one type of photonic topological insulator, whose realization only needs P-symmetry breaking. The domain wall between two valley-contrasting photonic crystals support robust edge states which can wrap around sharp corners without backscattering. Using the robust edge states, one can achieve the pulse transmission. Here, using time-domain measurement in the microwave regime, we show distortionless pulse transmission in a sharply bended waveguide. An Ω-shaped waveguide with four 120° bends is constructed with the domain wall between two valley photonic crystal slabs. Experimental results show the progress of Gaussian pulse transmission without distortion, and the full width at half maximum of the output signal was changed slightly in the Ω-shaped waveguide. By measuring steady state electric field distribution, we also confirmed the confined edge states without out-of-plane radiation which benefits from the dispersion below the light line. Our work provides a way for high-fidelity optical pulse signal transmission and develop high-performance optical elements such as photonic circuits or optical delay lines.**

# I. INTRODUCTION

Due to the important role that pulse transmission plays in modern optical science, people have studied pulse transmission in several photonic structures such as photonic ridge waveguides [1], photonic crystal waveguides [2-5], photonic crystal fibers [6, 7] and surface plasmon polariton waveguides [8-11]. Many optical phenomena such as slow light [3, 8], pulse reshaping [9] and third-harmonic generation [5] have been observed by measuring the pulse. Nevertheless, distortionless pulse transmission without scattering, resonance, and frequency dispersion in the waveguide is desirable in high-fidelity optical information transmission and processing.

Photonic crystal, as one type of artificial periodic optical structures, can realize several kinds of light manipulation, such as zero-index focusing [12, 13], slow light [14-16], cloaking [17]. Guiding light is one of the important applications of photonic crystal [18], which makes it possible to transmit optical information in integrated optoelectronic devices. However, when light encounter the defects, disorders or sharp bends, obvious backscattering will lead to energy loss and signal distortion. Therefore, distortionless pulse transmission without backscattering in photonic crystal is desirable. In the last decade, the introduction of topology into the optical systems has inspired the appearance of photonic topological insulators [19-21]. Robust transport of the edge states against defects has been achieved in gyromagnetic photonic crystals [22-24], coupled resonators [25, 26], helical photonic waveguides [27] or bianisotropic metacrystals [28]. It means that backscattering can be suppressed in photonic topological insulators, and they are expected to be used in high-fidelity optical information transmission. However, the employ of magnetism, time domain modulations or complicated design of metamaterials bring difficulties to the miniaturization for applying in communication regime. Recently, a new type of photonic topological insulator, i.e., valley photonic crystal (VPC) has been well explored and several devices have been demonstrated [29-36]. As the broken inversion symmetry is required, it can be applied within the all-dielectric structures [29, 34-36] and have the potential to be used in integrated optoelectronic devices [35, 36]. Domain wall between two different VPCs can guide light through sharp bends without backscattering. The waveguides constructed by this type of artificial dielectric structures are expected to guide pulse signals through sharp bends without distortion. The combination between robust transport and pulse signal transmission will be benefit for the potential application of VPC in the optical communication.

In this work, we designed a type of waveguide constructed by VPC slabs with dielectric rods standing on a perfect electric conductor (PEC) substrate. We experimentally measured and detailed analyzed the Gaussian pulse which is guided through an Ω-shaped waveguide, verified the distortionless pulse transmission against sharp bends. Field confinement of the valley-dependent edge states are also observed by measuring steady state electric field distributions.

## II. DESIGN OF WAVEGUIDE FOR DISTORTIONLESS PULSE TRANSMISSION

Distortionless transmission is a desirable feature to avoid the intersymbol interference in the information channel and becomes the key to guarantee the transmission bandwidth in optical communication. For a linear time-invariant system, it naturally means that different frequency components should propagate in the information channel without phase distortion and amplitude distortion, i.e. these components arrive at the same time with the same gain. This simultaneity requires that different frequency components should propagate with the same group velocity, which is estimated by group velocity dispersion (GVD) of the channel. And the same gain requires the different frequency components should suffer energy losses (scattering loss, radiation loss, absorption loss etc.) with a same rate, preferably no. Therefore, one would expect a waveguide channel with low GVD and low transmission loss.

VPC slab is an ideal platform to realize low-loss distortionless pulse transmission. The domain wall between two VPCs supports edge states due to their distinct topology. It can serve as a information channel. Since these edge states derive from the effective Hamiltonian near K and K' valleys, the dispersions are typically linear near these *k*-points. And this guarantees a similar group velocity (defined by $v_\text{g} = d\omega/dk$). The phase distortions caused by frequency-dependent time delays are thus mild. On the other hand, these topological edge states are robust against backscattering even when sharp bends exist, so that the in-plane scattering loss can be suppressed. Since the linear edge dispersion can locate below the light line by using appropriate parameters, the out-of-plane radiation loss is suppressed. Therefore, the amplitude distortion due to frequency-dependent loss can be minimized in our system. Besides, we use a PEC substrate to filter TE-like modes (*E* field is mostly parallel to the mirror plane of slabs, and is exactly parallel at *z* = 0) and leave only the TM-like (*E* field is mostly perpendicular to the mirror plane of slabs, and is exactly perpendicular at *z* = 0) edge states inside the complete bandgap [37]. Thus, the loss deriving from the cross coupling between TM-like and TE-like modes is reduced.

The two VPC slabs (VPC1 and VPC2) in our experiment are depicted in Fig. 1(a). Both of them are composed of ceramic rods ($\varepsilon_\text{r}$ = 9) arranged in a honeycomb lattice, which is a combination of two triangular sublattices (marked by A and B) with the same lattice constant of $a$ = 20 mm. The heights of all ceramic rods are $h$ = 17 mm. The diameter of rod A is $d_A$ = 7.5 mm ($d_A$ = 5.6 mm) while the diameter of rod B is $d_B$ = 5.6 mm ($d_B$ = 7.5 mm) for VPC1 (VPC2). VPC2 is the inversion partner of VPC1. Both slabs stand on an aluminum plate (assumed to be PEC in microwave regime). Figure 1(a) shows the TM-like bulk bands of two slabs, where VPC1 and VPC2 share a common complete bandgap. But their topologies are different due to the opposite valley Chern number $C_V = C_K - C_{K'}$ [35]. To see this, we calculated the Berry curvature of both VPC slabs [Fig. 1(c)]. As shown in the left panel of Fig. 1(c), the Berry curvature of the first bulk band of VPC1 is positive and concentrated around the K point in the momentum space (Berry curvature around the K' point is opposite to that around the K point due to the time reversal symmetry). On the contrary, the Berry curvature of VPC2 is negative and concentrated around the K point [right panel of Fig. 1(c)]. These opposite Berry curvature distributions between VPC1 and VPC2 show different topology and indicates the valley projected edge states. Note that as bulk bands of VPC slab are not well described by the Dirac equation, it leads to the deviation of the numerical valley Chern numbers from ±1. Detailed numerical Berry curvature and valley Chern number are investigated in Appendix A.

According to the bulk-edge correspondence, the domain wall between VPC1 and VPC2 (sketch in Fig. 1(b)) supports edge states and enables robust transport along the ΓK/ΓK' direction due to the suppression of intervalley scattering [35]. Figure 1(d) calculates the edge dispersion of the domain wall, where the green line highlights the edge states and the gray area covers the light cone. The experimental edge dispersion (obtained by Fourier transformation of the measured $E_z$ fields) labeled by the bright spots fits well with the simulated band curve. The frequency of the edge states below the light line ranges from 5.67 to 5.92 GHz, where the out-of-plane radiation should be neglectable. The calculated group velocity (defined as $v_g = \partial\omega/\partial k_x$) and GVD (defined as $GVD = (\omega^2/2\pi c)(1/v_g^3)(\partial v_g/\partial k_x)$) of the edge dispersion are shown in Figs. 1(e) and 1(f), respectively. Nearly constant group velocity (near-zero GVD) around 5.88 GHz (marked by black dashed lines) shows a good linearity of edge dispersion and guarantees the suppression of phase distortion. Therefore, we choose 5.88 GHz as the central frequency of the input Gaussian pulse in the experiments. The full bandwidth at half maximum (FBHM) of the pulse is set to be 40 MHz, i.e. frequency range 5.86 - 5.90 GHz (marked by the black boxes).

## III. EXPERIMENTAL DEMONSTRATION OF THE DISTORTIONLESS PULSE TRANSMISSION

In our experiments, pulse transmission through the waveguides will be studied in detail to confirm the distortionless pulse transmission. We employ an arbitrary waveform generator (Tektronix AWG70002A) to generate the Gaussian pulse. The input signal is launched into the waveguides from the entrance (i.e. the leftmost end of the channel) through a dipole antenna along the *z*-direction, and the $E_z$ field signal inside the waveguide channel is measured by another dipole antenna connected to an oscilloscope (Teledyne LeCroy MCM-Zi, 10-36Zi).

We first study the pulse transmission in a straight waveguide [Fig. 2(a)] to assess the phase distortion due to GVD. We monitor the evolution of the pulse waveform throughout the waveguide by measuring the time-domain signals at seven different positions (marked by blue and red stars, every $5a$ away from the entrance). Due to the insertion loss at the entrance, we only measure and compare the signals inside the waveguide channel. The signal received at the blue star (5$a$ away from the entrance) is shown in the inset of Fig. 2(a) and has a typical Gaussian shape. Its full width at half maximum (FWHM) is estimated as 23.52 ns by Gaussian fitting. Without loss of generality, we assume the arrival time at this position of the pulse center to be $t = 0$ and the distance of the position to be $d = 0$. Likewise, the time domain signals recorded at other positions (separately normalized by the maximum amplitude of themselves) are plotted in Fig. 2(c). We see that the pulse signals maintain a Gaussian shape during the transmission and no obvious deformation occurs. Their FWHMs (25.61 ns, 26.13 ns, 26.17 ns, 25.73 ns, 24.13 ns, 22.75 ns) and the arrival times of pulse center (2.69 ns, 4.84 ns, 6.56 ns, 7.52 ns, 7.31 ns, 8.62 ns) are obtained by performing Gaussian fitting and depicted in Fig. 2(c). We summarize the arrival times and FWHMs as a function of distance in Fig. 2(b). The arrival times have a relatively linear dependence on the distance, which means that the wave packet propagates almost with a constant speed. From Fig. 2b, it seems that the pulse arrives at the last two points (25$a$, 30$a$) in advance and earlier than the other points. This is because our measured signals are the superpositions of the forward pulse and the backward pulse reflected by the exit interface. And the reflected waves will affect the fitting pulse center. The averaged group velocity of the pulse center throughout the waveguide channel is estimated as $\bar{v}_\mathrm{g} = 0.247c$ using linear fitting (see the purple line in Fig. 2(b)), which agrees well with the theoretical prediction ($v_\mathrm{g} = 0.238c$, obtained in Fig. 1(d)). On the other hand, the FWHMs recorded at different positions are almost the same with about 11.27% variation throughout the waveguide channel, due to the similar group velocity of the different frequency components guaranteed by the linear edge dispersion. The little variation may be due to the superpositions of the forward pulse and the reflected pulse. Specifically, the FWHM of the pulse at the exit (the rightmost end of the channel, $d = 30a$) is 22.75 ns (only 3.27% deviated from the signal at $d = 0$), indicating little phase distortion.

Other than GVD, other frequency-dependent responses would also lead to waveform distortion, such as frequency-dependent scattering loss or frequency-dependent time delay caused by defects. Therefore, we further study an Ω-shaped waveguide with four 120° bends [Fig. 3(a)] and measure pulse transmission to assess the waveform distortion. Theoretically, the amplitude distortion caused by frequency-dependent scattering loss can be suppressed by the backscattering-immunity of topological edge states. Although these edge states would wrap around the corner eventually after a while, the time delays for different frequency components are generally different. This may result in phase distortion and will be studied in our experiment.

Similar to the straight waveguide, we monitor the waveform evolution at twelve different positions (marked by blue and red stars, $5a$, $10a$, $15a$, …, $50a$, $55a$, $59a$ away from the entrance along the channel). The signal received at the blue star ($5a$ away from the entrance) is shown in the inset of Fig. 3(a) and has a typical Gaussian shape. Its FWHM is estimated as 22.03 ns by Gaussian fitting. The arrival time of pulse center and the distance of the position are assumed to be zero. Likewise, the time domain signals recorded at other positions (separately normalized by the maximum amplitude of themselves) are plotted in Fig. 3(c). The pulse signals maintain a Gaussian shape and no obvious deformation occurs even after through four 120° sharp bends. It reflects that the frequency-dependency of time delays is not obvious in our system. Their FWHMs (22.04 ns, 21.05 ns, 21.64 ns, 22.33 ns, 20.91 ns, 21.42 ns, 23.66 ns, 24.96 ns, 21.04 ns, 22.11 ns, 21.00 ns) and the arrival times of pulse center (1.34 ns, 3.31 ns, 6.02 ns, 7.46 ns, 9.02 ns, 10.87 ns, 13.32 ns, 15.04 ns, 14.69 ns, 16.73 ns,

18.17 ns) are depicted in Fig. 3(c). We summarize the arrival times and FWHMs as a function of distance in Fig. 3(b). The relatively linear dependence of arrival times on the distance shows almost constant speed of the wave packet. The pulse arrives at the last three points ($45a$, $50a$, $54a$) in advance due to the backward pulse reflected by the exit interface. The averaged group velocity of the pulse center throughout the waveguide channel is estimated as $\bar{v}_g = 0.196c$ (see the purple line in Fig. 3(b)), which is 17.65% smaller compared with the theoretical prediction ($v_g = 0.238c$, obtained in Fig. 1(d)). This lower averaged group velocity is caused by the time delays at the corners. Although backscattering is suppressed, the pulse signal delays at the corners so that the averaged group velocity will be reduced. On the other hand, the FWHMs recorded at different positions are almost the same with only about 13.30% variation throughout the waveguide channel. The little variation may be due to superpositions of the forward pulse and the reflected backward pulse. Specifically, the FWHM of the pulse at the exit (the rightmost end of the channel, $d = 54a$) is 21.00 ns (only 4.68% deviation from the signal at $d = 0a$), indicating that time delays of different frequency component at the corners are almost the same. Distortionless pulse transmission in the Ω-shaped waveguide benefits not only from little phase distortion (guaranteed by low GVD and nearly frequency-independent time delay at the corners), but also from little amplitude distortion (guaranteed by the topologically-protected edge state).

Note that the total propagation time of pulse are different in two waveguides. Although their same sample size (the lengths along *x*-direction are both 35a), the optical distances of the two waveguides are different ($35a$ in straight waveguide, $59a$ in the Ω-shaped waveguide). Due to the different optical distances and the time delays at the corners, the propagation time in the Ω-shaped waveguide [18.17 ns] is longer than that in the straight waveguide [8.62 ns]. It provides a way to design on-chip robust optical delay line [25] with VPC slabs. The feature of nearly frequency-independent time delay of this delay line is useful for pulse signal transmission. In the experimental demonstration, we only considered the dispersionless part of edge states below the light line to avoid the propagation loss and pulse distortion. As a result, the used frequency bandwidth of pulse is only 4.3%, and it leads to the used long Gaussian signal. Thus, the frequency bandwidth the pulse transmission is narrower than that of conventional PC waveguide (~20% frequency bandwidth) and that of the conventional silicon strip waveguide (>30% frequency bandwidth). To broaden the frequency bandwidth (i.e., a shorter time duration) for pulse transmission, one can optimize the structural parameters of VPC waveguide by the boundary decoration (e.g., adjusting the position or radii of the rods nearest to the boundary). In Appendix B, we develop a method to increase the frequency bandwidth for pulse transmission and successfully increase the frequency bandwidth from 4.3% to 7.9%. It is important to note that although the frequency bandwidth in VPC waveguide is not so broadband, the pulse transmission is robust against sharp bends which is hard to realize in the conventional waveguide.

## IV. FIELD CONFINEMENT OF EDGE STATE

In previous sections, distortionless pulse transmission in VPC slab waveguides have been studied in time-domain. In this part, we discuss the field confinement of the edge state by measuring the steady state electric field distributions of the waveguides. A vector network analyzer (VNA, Keysight E5071C) is used to generate input source and measure the field distribution. A dipole antenna connected to VNA is used to excite the TM-like edge states, and another dipole antenna fixed on a motor scanning platform is used to receive the electric field $E_z$. We study the in-plane ($xy$-plane) field confinement first. We measure the steady state $E_z$ field distributions 1 mm above the straight and the Ω-shaped waveguides, the measured amplitudes at 5.88 GHz (center of the Gaussian pulse) are shown in Figs. 4a and 4b. Large amplitudes near the channel is observed not only in the straight waveguide, but also in the Ω-shaped waveguide. EM-wave does not scatter or radiate to the bulk crystal even when it propagates through four 120° sharp bends, showing good in-plane confinement. This is guaranteed by the backscattering immune property of VPCs. Further, we study the out-of-plane field confinement of the edge state (in $xz$-plane) in the straight waveguide. We predict that EM-wave in the frequency range of edge state below the light line (5.67 – 5.92 GHz) will be confined well in the waveguide for absence of the propagating mode in air. In Fig. 4(c), we show $E_z$ amplitude at 5.88 GHz in $xz$-plane (the cross-section along $x$-direction in the middle of the straight waveguide). White dashed line labels the top surface of the photonic crystal slab. The EM-wave is concentrated inside the waveguide, and decays rapidly when getting away from the top surface. This shows good out-of-plane confinement. Besides, we also confirm the confinement at different frequencies. For each frequency and $z$-coordinate, we sum up the measured $|E_z|$ of all $x$-positions, then the frequency spectrum of out-of-plane field confinement is shown in Fig. 4(d). We find that the $|E_z|$ in the frequency range of 5.67 – 5.92 GHz is larger and most of the fields are concentrated in the photonic crystal slab. While out of this frequency range, the localization of EM-wave becomes weak.

## V. CONCLUSIONS

In summary, we design a type of VPC slab waveguide standing on a PEC substrate. The distortionless pulse transmission in the waveguide is predicted by analyzing the linearity of edge dispersion and nontrivial band topology of the VPCs. Based on the time-domain measurement, we directly observe the distortionless Gaussian pulse signals in a straight waveguide and an Ω-shaped waveguide. By analyzing these pulse signals, we obtain their FWHMs. The results show a slight variation of the FWHMs. Besides, using the steady state near-field scan technology, we experimentally confirm field confinement of the edge states in the VPC waveguides. This work gives an experimental evidence of the distortionless pulse transmission against by sharp bends by combining the linear valley projected edge states under the light line. It is an interesting combination between robust transport and pulse transmission, which may have potential application of VPC in optical communication. The time-domain measurement also provides a different method to study the influence of topology on the pulse transmission. Taking advantage of a large time delay, the bended VPC waveguide has potential to be used as a robust optical delay line for pulse signal transmission in the integrated photonic circuits.

## ACKNOWLEDGEMENTS

This work was supported by National Natural Science Foundation of China (Grant Nos. 61775243, 11761161002, and 12074443), Natural Science Foundation of Guangdong Province (Grant No. 2018B030308005), Guangdong Basic and Applied Basic Research Foundation (Grant No. 2019B151502036), and Fundamental Research Funds for the Central Universities (Grant No. 20lgzd29).

## APPENDIX A: NUMERICAL BERRY CURVATURE AND VALLEY CHERN NUMBER

In this appendix, we study the numerical Berry curvature and valley Chern number of VPC slabs with different diameter difference between rods A and B [Fig. 5]. Based on the honeycomb lattice in Fig. 5(a), we gradually tune the difference between the diameters of rod A and B ($d_A$ and $d_B$), and study the evolution of Berry curvature around the K point and the numerical valley Chern number. In Fig. 5(b), when the diameters of rods A and B are slightly different ($d_A$ = 6.56 mm and $d_B$ = 6.54 mm), the Berry curvature is well localized near the K point, and the numerical valley Chern number $C_V = C_K - C_{K'} = 0.994 \approx 1$. When the diameter difference become larger (e.g., $d_A$ = 7.00 mm and $d_B$ = 6.10 mm), the localization of Berry curvature will become weaker [Fig. 5(c)]. The numerical valley Chern number is 0.709, which is smaller than 1. Figure 5(d) shows the case of VPC1 in the manuscript ($d_A$ = 7.50 mm and $d_B$ = 5.60 mm). The localization of Berry curvature is further weakened, and the valley Chern number is 0.464 which is much different from 1. Note that VPCs in Figs. 5(e), 5(f) and 5(g) are the inversion partner of those in Figs. 5(b), 5(c), and 5(d), so their numerical valley Chern number are respectively -0.994, -0.709, -0.464 which are the opposite values of valley Chern numbers of VPCs in Figs. 5(b)-5(d).

As a result, when the diameter difference between two rods becomes larger, the localization of Berry curvature will become weaker, and the numerical valley Chern number will increasingly deviate from ±1. This is because the ideal value of ±1 is derived by the Dirac equation of $H = v_D \left( \hat{\tau}_z \hat{\sigma}_x \delta k_x + \hat{\sigma}_y \delta k_y \right) + \lambda \hat{\tau}_z \hat{\sigma}_z$. However, when the diameter difference between two rods becomes larger, bulk bands of VPC are not well described by the Dirac equation, leading to the deviation of valley Chern numbers from ±1. Although valley Chern numbers of VPC1 and VPC2 deviate from ±1, Berry curvatures of them are opposite to each other and it indicates the valley projected edge states.

# APPENDIX B: ENLARGING FREQUENCY BANDWIDTH FOR PULSE TRANSMISSION

In this appendix, we develop a method to increase the frequency bandwidth of edge states for pulse transmission and increase the frequency bandwidth from 4.3% to 7.9%. To increase the frequency bandwidth, we can first reduce frequencies of bulk bands to make them locate below the light cone, then enlarge the bandgap, and lastly increase the ratio of edge states inside the bandgap. Based on the above ideas, we have the following optimization: First, to reduce frequencies of bulk bands, we can increase the effective refractive index of the VPC slab by enlarging the refractive index, diameter, or height of all rods. In the concrete example, we increase the relative permittivity of the all rods from 9 to 16. Then, to enlarge the bandgap, we further increase the diameter difference between the rod A and rod B. In the concrete example, the diameters of rod A and rod B in VPC1 (VPC2) are changed to $d_A$ = 8.00 mm and $d_B$ = 5.10 mm ($d_A$ = 5.10 mm and $d_B$ = 8.00 mm). Finally, to increase the ratio of edge states inside the bandgap, we can modify the boundary morphology. In the concrete example, we change the diameters of two rods at the middle of the boundary to 6.55 mm (name these two rods as transition layer in Fig. 2(a)). The introduction of this transition layer makes the edge states locate at the middle frequency of the bandgap, and it effectively increases the ratio of the edge states inside the bandgap. The dispersion of edge states of the optimized boundary is shown in Fig. 6(b). The frequency bandwidth is increased to 7.9% (4.73 – 5.12 GHz).

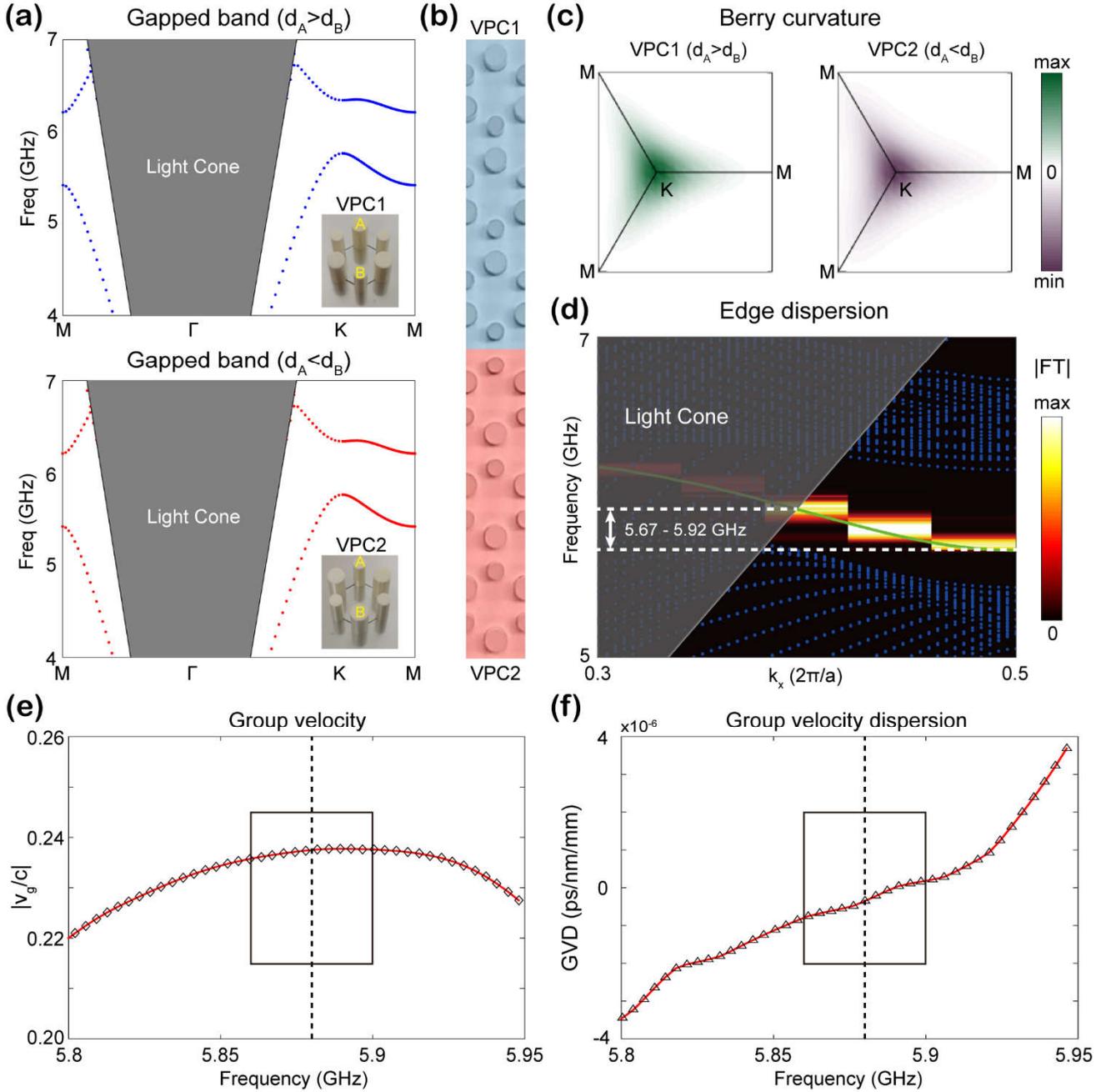

**Figure 1.** Group velocity ($v_g$) and group velocity dispersion (GVD) of the edge states. **(a)** The gapped bulk bands of two VPC slabs with the honeycomb lattice on a PEC substrate (at the top: VPC1 with $d_A > d_B$, at the bottom: VPC2 with $d_A < d_B$). The two slabs share the same bulk bandgap. **(b)** The domain wall between the two slabs. **(c)** Berry curvature of VPC1 and VPC2. **(d)** Edge dispersion of the domain wall in (b). The shadowed area is the light cone. Green solid curve corresponds to the simulated edge dispersion, bright spots plots experimental results. **(e)** Calculated group velocity ($v_g$) and **(f)** group velocity dispersion (GVD) of the edge states, which exhibits a uniform pulse velocity at the frequencies around 5.88 GHz. This guaranteed the distortionless transmission of the Gaussian pulse.

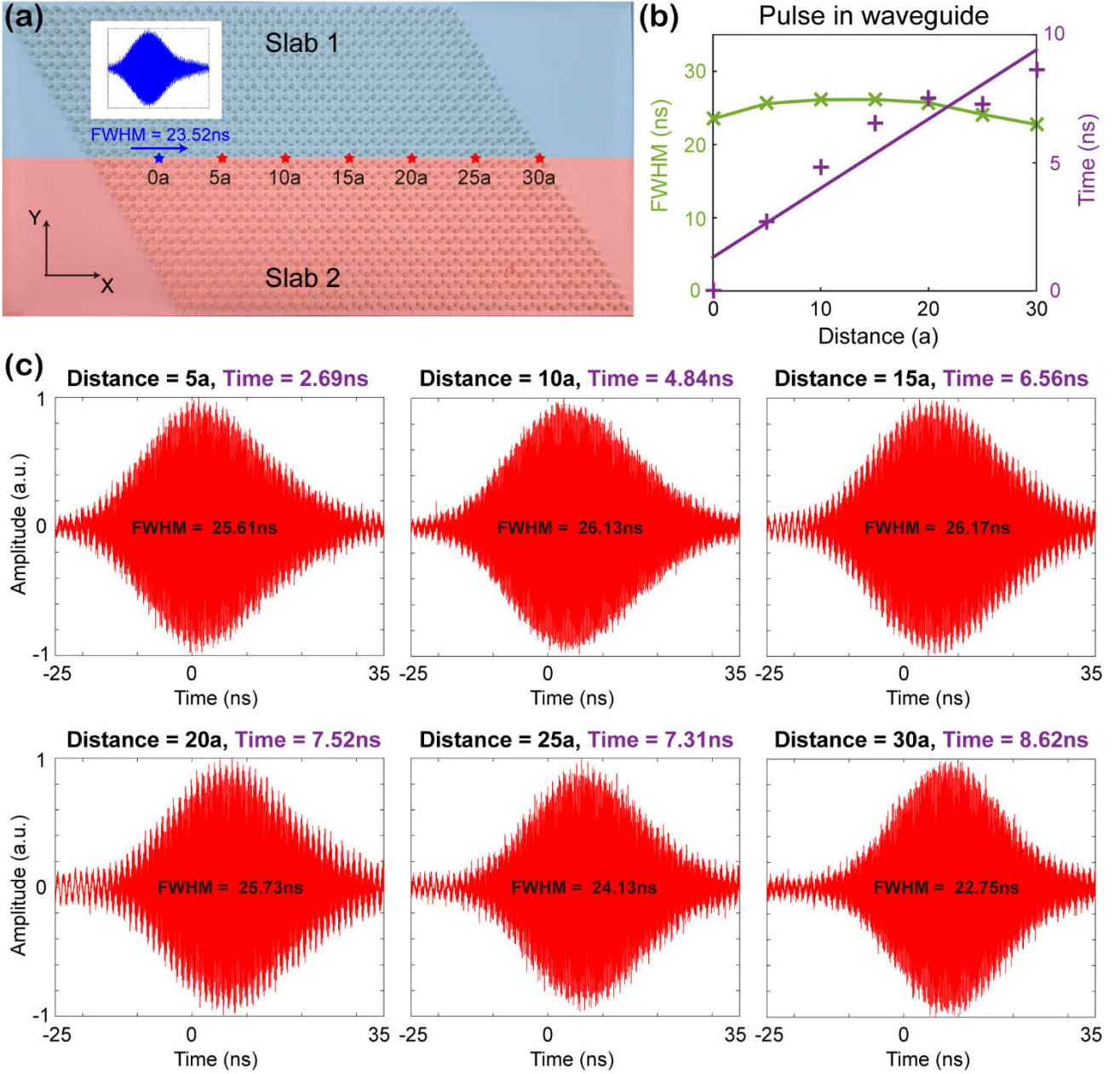

**Figure 2.** Distortionless pulse transmission in straight VPC slab waveguide. **(a)** Photo of the straight waveguide. It consists of two VPC slabs with opposite topological indices. Blue and red stars mark the positions of the detectors. Inset: measured Gaussian pulse which is centered at 5.88 GHz with a FBHM of 40 MHz. **(b)** Green cross: FWHM of the measured Gaussian pulse at different distances. FWHM of the pulse at the rightmost exit changes only 3.27% compared with that of the input pulse. Purple cross: the arrival time of pulse center as a function of distance. Purple line: The linear fitting of relationship between arrival time and distance, indicating that the averaged group velocity of the pulse throughout the waveguide is $\bar{v}_g = 0.247c$. **(c)** Measured pulse at different distances (marked by red stars in (a)) in the straight waveguide.

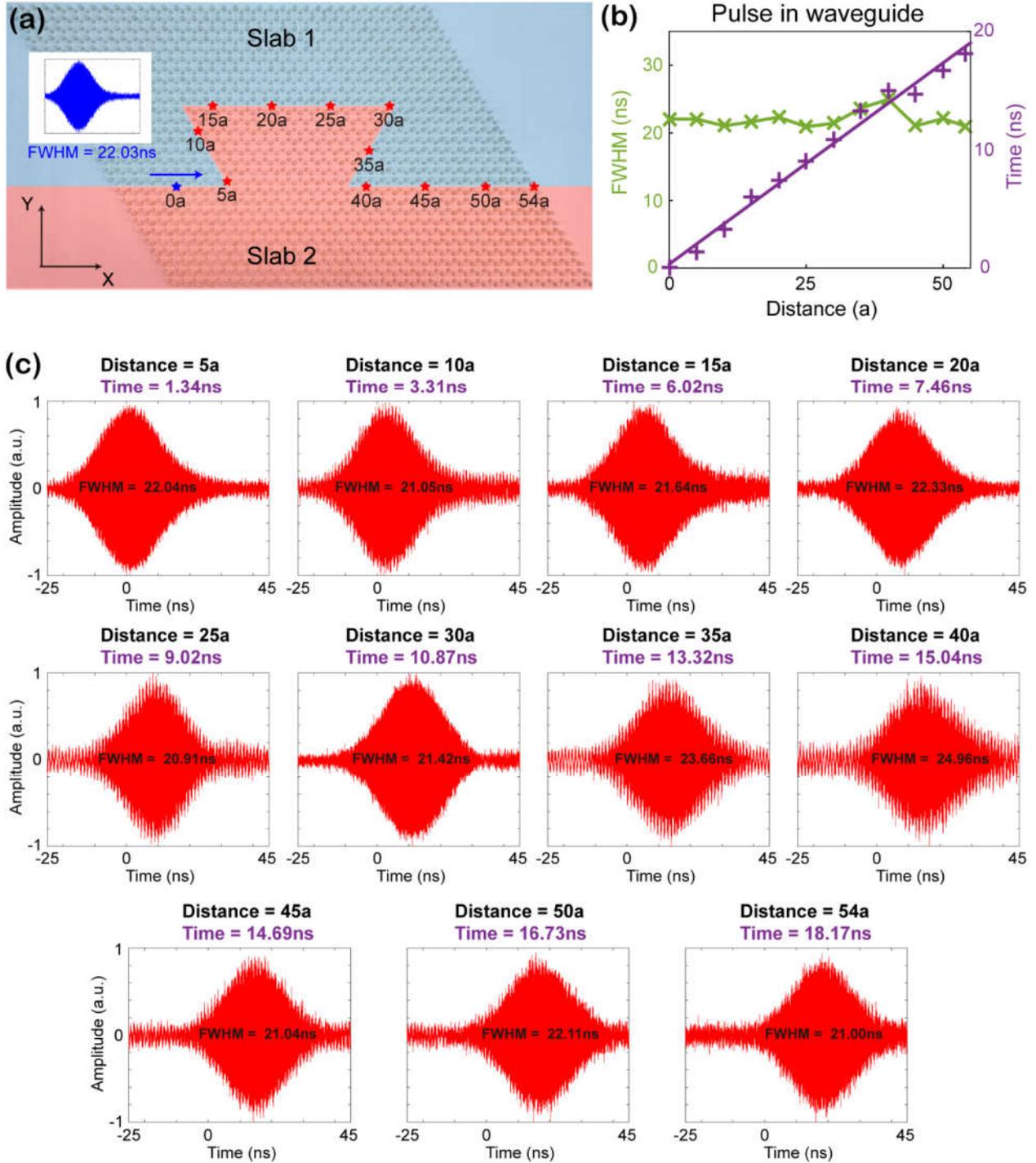

**Figure 3.** Distortionless pulse transmission in an Ω-shaped VPC slab waveguide. **(a)** Photo of the Ω-shaped waveguide. Blue and red stars: positions of the detectors. Inset: measured Gaussian pulse (centered around 5.88 GHz, FBHM = 40 MHz) at the blue star. **(b)** Green cross: FWHM of the measured Gaussian pulse at different distances. FWHM of the pulse at the rightmost exit changes only 4.68% compared with the input pulse. Purple cross: the arrival time of pulse center as a function of distance. Purple line: linear fitting of the arrival time, indicating that the averaged group velocity of the pulse throughout the waveguide is $\bar{v}_g = 0.196c$. **(c)** Measured pulse at different distances (marked by red stars in (a)) in the Ω-shaped waveguide.

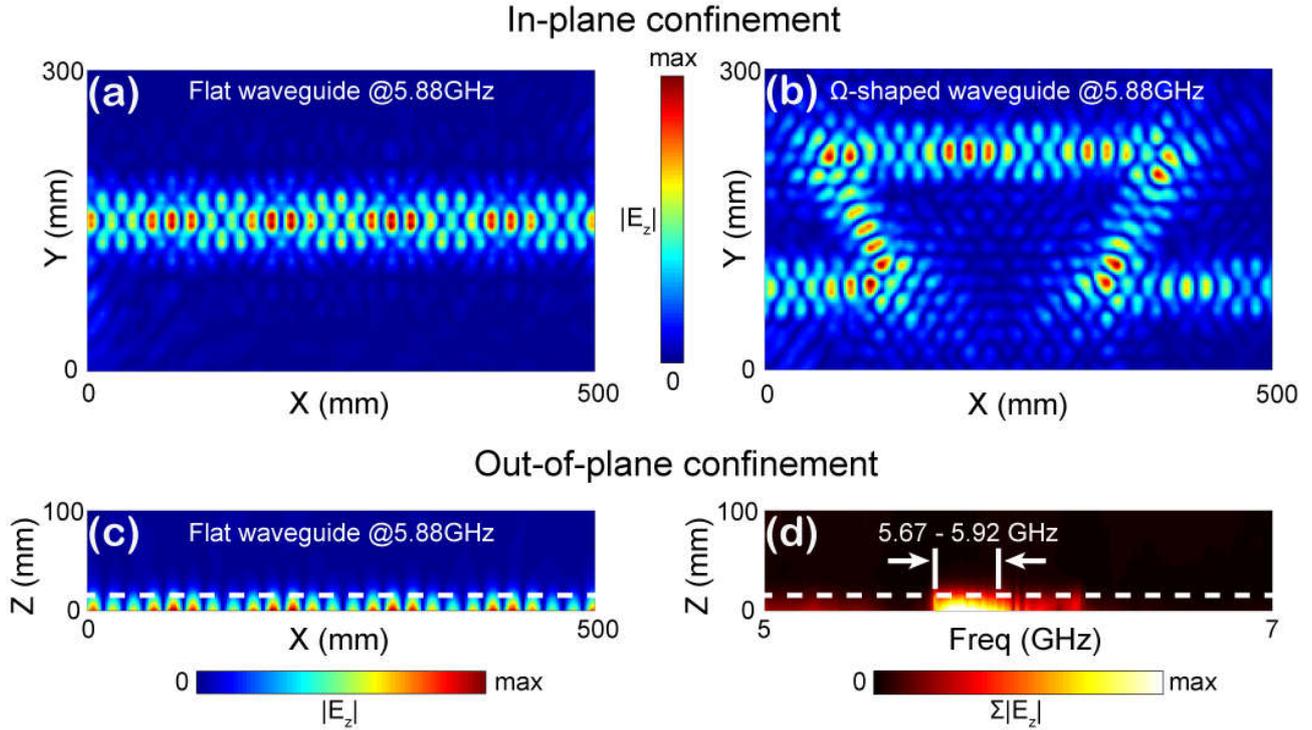

**Figure 4.** Field confinement of the measured edge states. **(a-b)** In-plane confinement. Measured $E_z$ field distributions of the edge states in **(a)** the straight and **(b)** the Ω-shaped waveguide at 5.88 GHz (center of the Gaussian pulse) in the xy-plane, confirmed the good confinement and robust transport of the edge states. **(c-d)** Out-of-plane confinement. **(c)** Measured $E_z$ field distributions in the straight waveguide at 5.88 GHz in the xz-plane. Out-of-plane distribution shows that the edge states are confined well in the waveguide, but decay rapidly when getting away from the top surface of the sample (denoted by white dashed line). **(d)** Field confinement of the edge states for different frequencies.

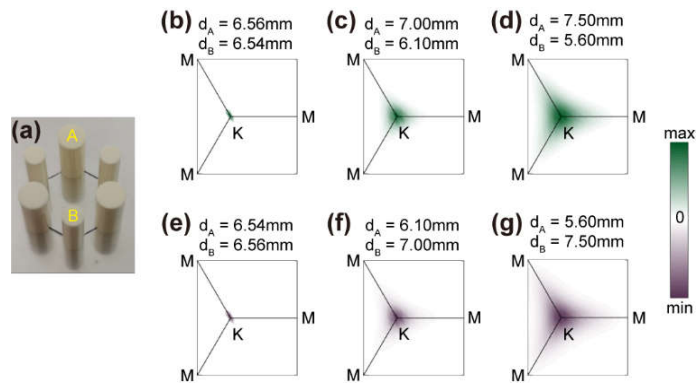

**Figure 5.** Calculated Berry curvatures of VPC slabs of two rods with different diameters. **(a)** Photo of the unit cell of VPC slab with honeycomb lattice. **(b-d)** Berry curvatures around the K point when $d_A > d_B$. **(e-g)** Berry curvatures around the K point when $d_A < d_B$.

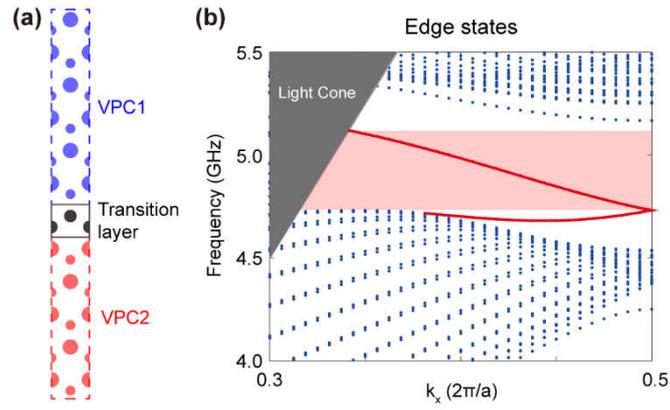

**Figure 6.** A design for edge states with a larger frequency bandwidth. **(a)** The domain wall with a transition layer between two VPC slabs. **(b)** Edge dispersion of the structure in (a). The frequency bandwidth of edge states for pulse transmission is 7.9% (4.73 – 5.12 GHz).